\documentclass[epj, twocolumn]{svjour}
\usepackage{latexsym}
\usepackage{graphics}
\usepackage{amsmath}
\newcommand{\ba}{\vec{\bf a}}
\newcommand{\CS}{Ca$_{2-x}$Sr$_x$RuO$_4$}
\newcommand{\Sr}{Sr$_2$RuO$_4$}

\usepackage{epstopdf}
\usepackage{amssymb}
\usepackage{epsfig}

\date{Received: \today  / Revised version: date}

\begin{document}

\title{Interplay of metamagnetic and structural transitions in Ca$_{2-x}$Sr$_x$RuO$_4$.}

\author{R. Rold\'{a}n\inst{1,2}, A. R\"{u}egg\inst{3} and M. Sigrist\inst{3}}
\institute{ Laboratoire de Physique des Solides, Univ. Paris-Sud, CNRS, UMR 8502, F-91405 Orsay Cedex, France. \and Instituto de Ciencia de Materiales de Madrid, CSIC, Cantoblanco, E-28049 Madrid, Spain 
\and Theoretische Physik, ETH Z\"urich, CH-8093 Z\"urich, Switzerland}

\abstract{
Metamagnetism in layered ruthenates has been interpreted as a novel kind of quantum critical behavior. 
In an external magnetic field, \CS~ undergoes a metamagnetic
transition accompanied by a pronounced magnetostriction effect. In
this paper we present a mean-field study for a microscopic model
that naturally reproduces the key features of this
system. The phase diagram calculated is equivalent to the
experimental $T$-$x$ phase diagram. The presented model also gives
a good basis to discuss the critical metamagnetic behavior
measured in the system.
\PACS{75.30.-m,75.50.-y,75.80.+q
     } 
}
\maketitle

\section{Introduction}

\CS~ has attracted interest for its complex and puzzling phase diagram including
metallic as well as Mott-insulating magnetic phases \cite{N00} which depend in a subtle way 
on structural properties. 
Recent magnetostriction experiments \cite{B05,B06} in connection with the metamagnetic transition (MMT) for $ x \approx 0.2 $ underline the strong coupling of the electronic properties to the lattice. 
This may be taken as a hint for the relevance of localized electronic orbital and spin degrees of freedom
as  can be found in a number of transition metal oxides. 
In particular, the mutual influence of spin and orbital
degrees of freedom plays an important role in the behavior of
manganites, ruthenates or titanates \cite{Science-Nagaosa}.

\begin{figure}[t]
{\includegraphics[width=0.8\linewidth]{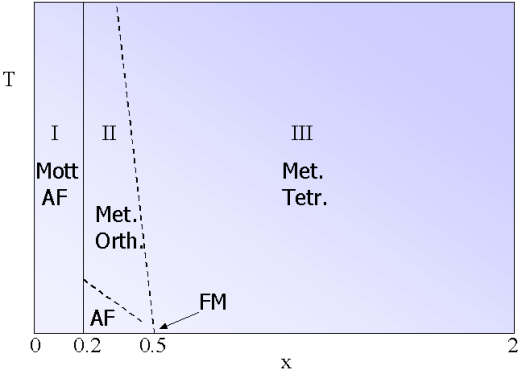}} \caption{Sketch of
the temperature-doping phase diagram of Ca$_{2-x}$Sr$_x$RuO$_4$.
Region I corresponds to a Mott insulator with long range
antiferromagnetic order. Region II is characterized by a metallic
behavior with orthorhombic lattice symmetry and antiferromagnetic
correlations at low temperatures. Region III corresponds to a
paramagnetic metal with tetragonal symmetry in the lattice. For
$x\rightarrow 0.5$ there are strong ferromagnetic correlations at
low temperatures.} \label{fig:PD}
\end{figure}

The result of the apparent interplay between orbital and magnetic
correlations in \CS can be seen in the complex $T$-$x$ phase
diagram shown in Fig.~\ref{fig:PD}. Both Sr$^{2+}$ and Ca$^{2+}$ are isovalent ions so that the substitution
of one by the other does not change the number of conductance electrons.
Sr$_2$RuO$_4$ is a good metal and the \emph{a priori} expected change due to
the substitution of Sr by Ca should be increasing metallicity, because the
doping with a smaller ion (Ca)  would imply a widening of the
band. This is not the case because the smaller Ca-ion induces lattice distortions which alter
the overlap integrals and the crystal fields of the relevant electronic orbitals. 
Actually, Ca$_2$RuO$_4$ behaves as a Mott insulator, and  the
evolution between these end-members builds a very rich phase diagram where
different structural and magnetic phases appear (Fig.\ref{fig:PD}). For $x=2$, corresponding to \Sr, the system has
tetragonal symmetry with the RuO$_6$-octahedra slightly elongated along the
$c$-axis. Ca substitution initially induces the rotation of RuO$_6$ octahedra around the c-axis in order to accommodate the smaller ions. The system behaves still as a paramagnetic metal with tetragonal
symmetry. 
Further doping with Ca reduces the conductivity and
the susceptibility increases when approaching $x\rightarrow 0.5$,
becoming a Curie-like susceptibility \cite{NM00}. At $x=0.5$ there is a structural phase transition, where the crystallographic structure of the \CS~ series changes from tetragonal to orthorhombic through a second-order phase transition. For $x<0.5$ there is, besides the c-axis rotation, a tilting of the
RuO$_6$ octahedra leading to a reduction of the symmetry and a reduction of the c-axis lattice constant. Moreover, these distortions are responsible for a narrowing of the conduction bands which in turn enhances the correlation effects. In the region $0.2\leq x \leq 0.5$ the experiments suggest the appearance of a low-temperature antiferromagnetic (AFM) order. Furthermore, the application of a magnetic field leads to a metamagnetic transition which influences the tilting of the
RuO$_6$-octahedra \cite{B05,B06}. Within the region $x\leq 0.2$ the system is
a Mott-insulator with true long-range AFM order and a total
spin of $S=1$.

In general, a MMT can occur for a material which, under the
application of an external field, undergoes a first order
transition or is close to the critical endpoint of such a transition 
to a phase with strong ferromagnetic correlations. This is
experimentally observed by a very rapid increase of the magnetization
over a narrow range of applied magnetic field. The field dependence
of the magnetization and magnetoresistance of \CS~ has been analyzed in
Ref.~\cite{N03}. There, a MMT to a
highly polarized state with a local moment of $S=1/2$ is found. These
measurements are interestingly supplemented by the magnetostriction
experiments published in Ref.~\cite{B05,B06} where the
change of the lattice constants as a function of the magnetic field
for different temperatures is shown.  
Namely, these results demonstrate that
crossing the MMT leads to 
an elongation of the $c$ axis accompanied by a shrinking
along both in-plane directions. Apparently, the application of a high
magnetic field at low temperatures can reverse the structural distortion that
occurs for $0.2\leq x \leq 0.5$ upon cooling in zero field. 
Furthermore, apart from different energy scales, the qualitative effects associated with the MMT are independent of the magnetic field direction.

All the members of the \CS~ family have 4 electrons in the $t_{2g}$ orbitals
of the Ru 4$d$ shell. What is different is the occupation of these orbitals in
the two end-members of the phase diagram: while for $x=0$ there is an average
occupation of two electrons in the $d_{xy}$ orbital and the other two in the
$d_{yz}$ and $d_{zx}$ orbitals, for $x=2$ there is a fractional occupation of
4/3 in the $d_{xy}$ band and 8/3 in the $d_{yz}$-$d_{zx}$-bands.   LDA calculations for this concentration give three Fermi surface
sheets, one with essentially $xy$ and two with mixed $\{ xz, yz\}$ character
\cite{O95}. The first one is usually called $\gamma$ band while the others are
labeled by $\alpha$ and $\beta$ bands. For intermediate values of $x$ the orbital occupation is still a matter of debate \cite{ANKRS02,FANG04,OKA04,KO07,LIEB07}.

In this work we focus on the region II and the Ca-rich part of region III in the schematic $T-x$ phase diagram. For the microscopic description we follow the scenario of an orbital-selective Mott insulator. This scenario was put forward by Anisimov \emph{et al.} \cite{ANKRS02} in order to explain the unexpected effective magnetic moment close to a $S=1/2$ spin for $0.2 \lesssim x \lesssim 1.5$ \cite{N03}. 
Assuming that the orbital occupation in this region is $(n_{\alpha,\beta},n_{\gamma})\approx(3,1)$, they proposed that the electrons in the $\{\alpha,\beta\}$ bands undergo an orbital-selective Mott transition (OSMT) while the $\gamma$ band remains metallic. In this scenario the experimental observation of the $1/2$ effective spin is assigned to the localized hole in the $\{\alpha,\beta\}$ bands. Angular magnetoresistance oscillations measurements \cite{BS05} indeed show a strong dependence of the Fermi surface on the Ca concentration which is consistent with the scenario of coexisting itinerant and localized $d$-electronic states. Furthermore, from the theoretical point of view, there is by now a consensus that an OSMT can in principle occur in multi-band Hubbard models under rather general conditions \cite{Liebsch:03,Koga:04,Ruegg:05,Ferrero:05,Medici:05,Arita:05,Knecht:05,Costi:07,Inaba:06}. However, it is still unclear to which extend this concept is applicable in the \CS~ system \cite{Lee:2006,Wang:2004}. Despite of these uncertainties, there is little doubt that the localized degrees of freedom play an important role for the understanding of the puzzling physics of this material, and in the following we will assume that the concept of the OSMT is valid as a lowest order picture for Ca concentrations corresponding to region II and at the boundary of region III of the $T-x$ phase diagram.

Using a mean-field description we focus on the interplay between structural distortion and magnetic and orbital ordering. We find a theoretical phase diagram which can be related to the experimental one. Furthermore, our calculations qualitatively reproduce the metamagnetic transition accompanied by a structural transition observed in the system. The paper is organized as follows: in Sec.~\ref{sec:model} we introduce the microscopic model. A mean-field analysis is performed in Sec.~\ref{sec:mf}. The results for zero and finite magnetic field are given in Sec.~\ref{sec:res}. In Sec.~\ref{sec:exp} we relate our results to the experimental measurements. We summarize the main conclusions of this work in Sec.~\ref{sec:con}.

\section{The model}
\label{sec:model}
Following the scenario of the OSMT we assume one hole in the \{$\alpha,\beta$\} bands and focus on the localized orbital and spin degrees of freedom \cite{ANKRS02,ST04}. Neglecting for the our discussion the itinerant $\gamma$ band it is natural to consider a two-dimensional extended Hubbard model of the form

\begin{eqnarray}
\label{ExtendedH}
{\cal H}_{\alpha,\beta}& = &-t \sum_{i,{\ba},s}
\left(c_{i+a_y,yz, s}^{\dag} c_{i,yz,s} + c_{i+a_x,zx,s}^{\dag}
c_{i,zx,s} + h.c.\right)  \nonumber\\
&& - \mu \sum_{i,s,\nu}
c_{i,\nu,s}^{\dag} c_{i,\nu,s} \nonumber\\
&&+ U
\sum_{i} \sum_{\nu} n_{i\nu \uparrow} n_{i \nu \downarrow} + U'
\sum_{i} n_{i,zx} n_{i,yz}\nonumber \\
 & &+J_H\sum_{i,s,s'} c_{i,yz,s}^{\dag} c_{i,zx,s'}^{\dag} c_{i,zx,s}
c_{i,yz,s'}
\end{eqnarray}
where $ c_{i,\nu,s}^{\dag} $ ($ c_{i,\nu,s} $) creates (annihilates) an
electron on site $ i $ with orbital index $ \nu $ ($ = yz,zx $) and
spin $ s $ ($ n_{i,\nu,s} = c_{i,\nu,s}^{\dag} c_{i,\nu,s} $, $
n_{i,\nu} = n_{i, \nu, \uparrow} + n_{i, \nu, \downarrow} $; $ {\ba}
= (a_x , a_y ) = (1,0) $ or $ (0,1) $ basis lattice vector). With
this Hamiltonian we restrict ourselves to nearest-neighbor hopping
and on-site interaction for the intra- and inter-orbital Coulomb
repulsion, $U$ and $U'$, respectively, and the Hund's rule coupling
$J_H$. The hopping terms considered in this model come from the
$\pi$-hybridization between the Ru-$d$ and O-$p$-orbitals and lead
to the formation of two independent quasi-one-dimensional bands: the
band associated to the $d_{yz}$-orbital disperses only in the
$y$-direction while the band associated to the $d_{zx}$-orbital
disperses in the $x$-direction.

In the strongly interacting limit it was proposed that the $ \alpha $-$\beta$-bands absorb 3 of the four electrons available per site and form a Mott-insulating state with localized degrees of freedom, spin 1/2 and orbital  \cite{ANKRS02}.  The local orbital degree of freedom can be represented as an isospin
configuration $ | + \rangle $ and $ | - \rangle $ corresponding to the
singly occupied $ d_{zx} $ and $ d_{yz} $ orbitals, respectively. The isospin operators therefore may be defined as:
\[
I^z | \pm \rangle = \pm \frac{1}{2} | \pm \rangle, \qquad I^+ |-
\rangle = | + \rangle, \qquad I^- | + \rangle = | - \rangle \; .
\]
Taking into account the additional spin 1/2 degree of freedom  ($ |\uparrow \rangle $ and
$|\downarrow \rangle$) leads to four possible
configurations at each site, represented by the states 
\[
\{ | \uparrow + \rangle, \; | \uparrow - \rangle , \; | \downarrow +
\rangle , \; |\downarrow - \rangle \}.
\]
Within second order perturbation in $t/U$ it is possible to derive from ${\cal H}_{\alpha,\beta}$ an effective model describing the interaction between the localized degrees of freedom. One finds the following Kugel-Khomskii-type
model \cite{ANKRS02}:
\begin{eqnarray}\label{Heff}
{\cal H}_{eff}  =  J \sum_{i,{\ba}} &&\Big\{ \left[A
(I^z_{i+{\ba}} +
    \eta_{{\ba}})(I^z_{i} +
\eta_{{\ba}}) +B \right]
{\bf S}_{i+{\ba}} \cdot {\bf S}_{i} \nonumber \\
& & + [C (I^z_{i+{\ba}} + \eta'_{{\ba}})(I^z_i +
\eta'_{{\ba}}) + D] \Big\}
\end{eqnarray}
where $J=4t^2/U$. We have imposed the approximatively valid
relation $U=U^{\prime}+2J_H$ and have assumed that $\alpha=U^{\prime}/U>1/3$. The parameters $A,B,C,D,\eta_{\ba}$ and $\eta_{\ba}^{\prime}$ are functions of $\alpha$ alone and have been given elsewhere \cite{ANKRS02,ST04}. The energy scale $JC>0$ of the isospin
coupling is the largest in the present Hamiltonian. Therefore, in a mean-field approximation, one expects antiferro-orbital (AFO) order below a critical temperature $T_{AFO}\sim JC$ ($C>0$). On the other hand, the value of the spin-spin
interaction depends on the orbital order and lies between $J_1=J[A(\eta_{\mathbf{a}}^2-1/4)+B]<0$
and $J_2=J[A\eta_{\mathbf{a}}^2+B]$. Thus, in the presence of AFO order the spin will align
ferromagnetically (FM) below a critical temperature
$T_{FM}\sim-J_1$. If, however, AFO order is suppressed, as in the
case of an orthorhombic distortion (see below) the spin-spin
coupling is given by $J_2$. We mention here that the sign of $J_2$
depends on the value of $\alpha$. In particular, $J_2<0$ for $\alpha<\alpha_c=0.535$ and consequently we expect FM order at low temperatures whereas
for $\alpha>\alpha_c$ we have $J_2>0$ and antiferromagnetic (AFM)
order sets in at sufficiently low temperatures. To be consistent with experiments we will choose throughout this article a value $\alpha=0.75$.

The Sr substitution for Ca acts as an effective negative pressure. In order to account for the orthorhombic distortion due to the tilting of the
RuO$_6$ octahedra, we introduce a new term in the Hamiltonian ${\cal
  H}_{dist}$, defined as
\begin{equation} {\cal
    H}_{dist}=\frac{1}{2}GN(\varepsilon-\varepsilon_0)^2+K\varepsilon\sum_iI_i^x,
\label{eq:Hdis}
\end{equation}
where $G$ is the elastic constant, $N$ is the number of Ru atoms and $K$ is a coupling constant. $\varepsilon$ is a strain-field which accounts for an orthorombic
distortion. Any orthorombic distortion
yields a uniform bias for the local orbital configuration which suppresses AFO
ordering. This is modeled by the coupling of $\varepsilon$ to the orbital degrees of freedom. In other words, the orthorombic distortion introduces a transverse
field which aims to align the isospins. 
The first term in Eq.~(\ref{eq:Hdis}) is a measure of the lattice elastic
energy. In addition to the strain $\varepsilon$ driven by orbital correlations we assume a constant contribution $\varepsilon_0$. We do not specify further the origin of this contribution but it might include effects of the $\gamma$ band or other, non-electronic, mechanisms. For actual calculations we fix the value at $\varepsilon_0=0.1$. We will discuss later to which extend we can relate the elasticity $G$ in the theoretical model to the Sr concentration $x$ in \CS.

Finally, in order to study the metamagnetic transition, we introduce a
coupling of the system to a magnetic field, by the inclusion of
the term ${\cal H}_{mag}$,
\begin{equation}\label{Hmag}
{\cal H}_{mag}=-g\mu_BH\sum_iS_i^x,
\end{equation}
where $g$ is the electron gyromagnetic factor, $\mu_B$ is the Bohr
magneton, $\mu_B=\frac{e\hbar}{2m_e}$ and $H$ is the magnetic field strength. With all these
ingredients, the full Hamiltonian can be written as

\begin{equation}\label{FullHamil}
{\cal H}={\cal H}_{eff}+{\cal H}_{dist}+{\cal H}_{mag}.
\end{equation}
In the next section we treat this model in a mean-field approximation.

\section{Mean-field analysis.}
\label{sec:mf}
The mean-field decoupling for ${\cal H}_{eff}$
reproduces well some of the experimentally observed features of
\CS~ in the $x$ region where
the band filling corresponds to the $(n_{(\alpha,\beta)},n_{\gamma})=(3,1)$ orbital occupation, as shown in Ref. \cite{ANKRS02}. Here we extend this analysis to the full Hamiltonian
Eq.~(\ref{FullHamil}) and obtain the zero and finite magnetic field
phase diagrams where the different competing orders of the system
are represented. In the presence of  a transverse magnetic field, there appears an uniform component of the magnetization in the
$x$-direction, 
\begin{equation}
\langle S_i^x \rangle=m_0,
\end{equation}
as well as a staggered component in the $z$-direction, 
\begin{equation}
\langle S_i^z \rangle=\left\{
\begin{array}{ccc}
  m_s &\,\,\,\, \mathrm{if} &\,\,\,\, i\in A; \\
  -m_s &\,\,\,\, \mathrm{if} &\,\,\,\, i \in B. \\
\end{array}
 \right.
\end{equation}
Here, we have made use of the bipartite structure of the Hamiltonian Eq.~(\ref{FullHamil}): $A$ and $B$ label the two sublattices. In addition, we introduce the staggered isospin component in the $z$-direction
\begin{equation}
\langle I_i^z \rangle=\left\{
\begin{array}{ccc}
  t_s &\,\,\,\, \mathrm{if} &\,\,\,\, i\in A; \\
  -t_s &\,\,\,\, \mathrm{if} &\,\,\,\, i \in B. \\
\end{array}
 \right.
\end{equation}

The partition function of the system can be calculated by ${\cal
Z}(\beta)=\mathrm{Tr} e^{-\beta {\cal H}}$, where $\beta=1/k_BT$. On the mean-field level, the
bipartite nature of the lattice splits the system into two subsystems, so we can express the partition function as
\begin{equation}
{\cal Z}(\beta)=\left(e^{-\beta E_0}\right)^N\prod_{\substack{i\in A,B \\ \alpha \in t,s}}
{\cal Z}_i^{\alpha}
\end{equation}
where $E_0$ denotes the energy density (per hole) corresponding to
the term in the mean-field Hamiltonian that does not couple to any
spin or isospin operator,
\begin{eqnarray}
E_0&=&J\Big[6A\,t_s^2(m_0^2-m_s^2)-2(A\,\eta_{\ba}^2+B)(m_0^2-m_s^2)\nonumber\\
&&+2C\,t_s^2+2C\,{\eta_{\ba}^{\prime}}^2+2D\Big]+\frac{G}{2}(\varepsilon-\varepsilon_0)^2.
\end{eqnarray}
$N=N_A+N_B$ is the total number of sites.
Now we can define the orbital ${\cal Z}_i^t$ and magnetic ${\cal
Z}_i^s$ one-particle partition functions as \cite{BZ74}:
\begin{eqnarray}
{\cal
Z}_{i\in A,B}^t&=&\mathrm{Tr}_ie^{-\beta \vec{H}_i^t\cdot\vec{I}_i}\\
{\cal Z}_{i \in A,B}^s&=&\mathrm{Tr}_ie^{-\beta \vec{H}_i^s\cdot\vec{S}_i}
\end{eqnarray}
where $\vec{H}_i^t$ and $\vec{H}_i^s$ are the molecular
field vectors that couple to the isospin and spin degrees of
freedom. If we denote the total free energy of the system by $F$,
the free energy per site (per hole) ${\cal F}=F/N=-\ln {\cal Z}/\beta N$ is
\begin{equation}
{\cal F}=E_0-\frac{1}{2\beta}\left(\ln
{\cal Z}_A^t +\ln {\cal Z}_B^t+\ln {\cal Z}_A^s+\ln {\cal
Z}_B^s\right)
\end{equation}
Eventually, the free
energy of the system for the given mean fields is found to be
\begin{eqnarray}\label{Fh}
{\cal
F}&=&J\Big[6A\,t_s^2(m_0^2-m_s^2)-2(A\,\eta_{\ba}^2+B)(m_0^2-m_s^2)\nonumber\\
&&+2C\,t_s^2+2C\,{\eta_{\ba}^{\prime}}^2+2D\Big]+\frac{G}{2}(\varepsilon-\varepsilon_0)^2\nonumber\\
&&-\frac{1}{\beta}\ln\left[2\cosh\left(\frac{\beta}{2}\sqrt{f_1}\right)\right]\nonumber\\
&&-\frac{1}{\beta}\ln\left[2\cosh\left(\frac{\beta}{2}\sqrt{f_2}\right)\right],
\end{eqnarray}
where
\begin{eqnarray}
f_1&=&K^2\varepsilon^2+16J^2\left[A\,t_s(m_0^2-m_s^2)+C\,t_s\right]^2,\nonumber\\
f_2&=&\left\{-g\mu_BH+4Jm_0\left[A(\eta_{\ba}^2-t_s^2)+B\right]m_0\right\}^2\nonumber\\
&&+16J^2m_s^2\left[A(\eta_{\ba}^2-t_s^2)+B\right]^2.
\end{eqnarray}
The values of the mean fields are determined by the solution of the self-consistency equations 
\begin{equation}
0=\frac{\partial\cal{F}}{\partial\phi}
\end{equation}
where $\phi=\varepsilon, t_s, m_0, m_s$. These equations are solved numerically. Because
\begin{equation}\label{I^x}
\langle I^x\rangle=\frac{G}{K}(\varepsilon-\varepsilon_0)
\end{equation}
the ferro-orbital order parameter is directly related to the difference $\varepsilon-\varepsilon_0$.

\section{Results}\label{sec:res}
\begin{figure}[t]
  \includegraphics[width=0.44\linewidth]{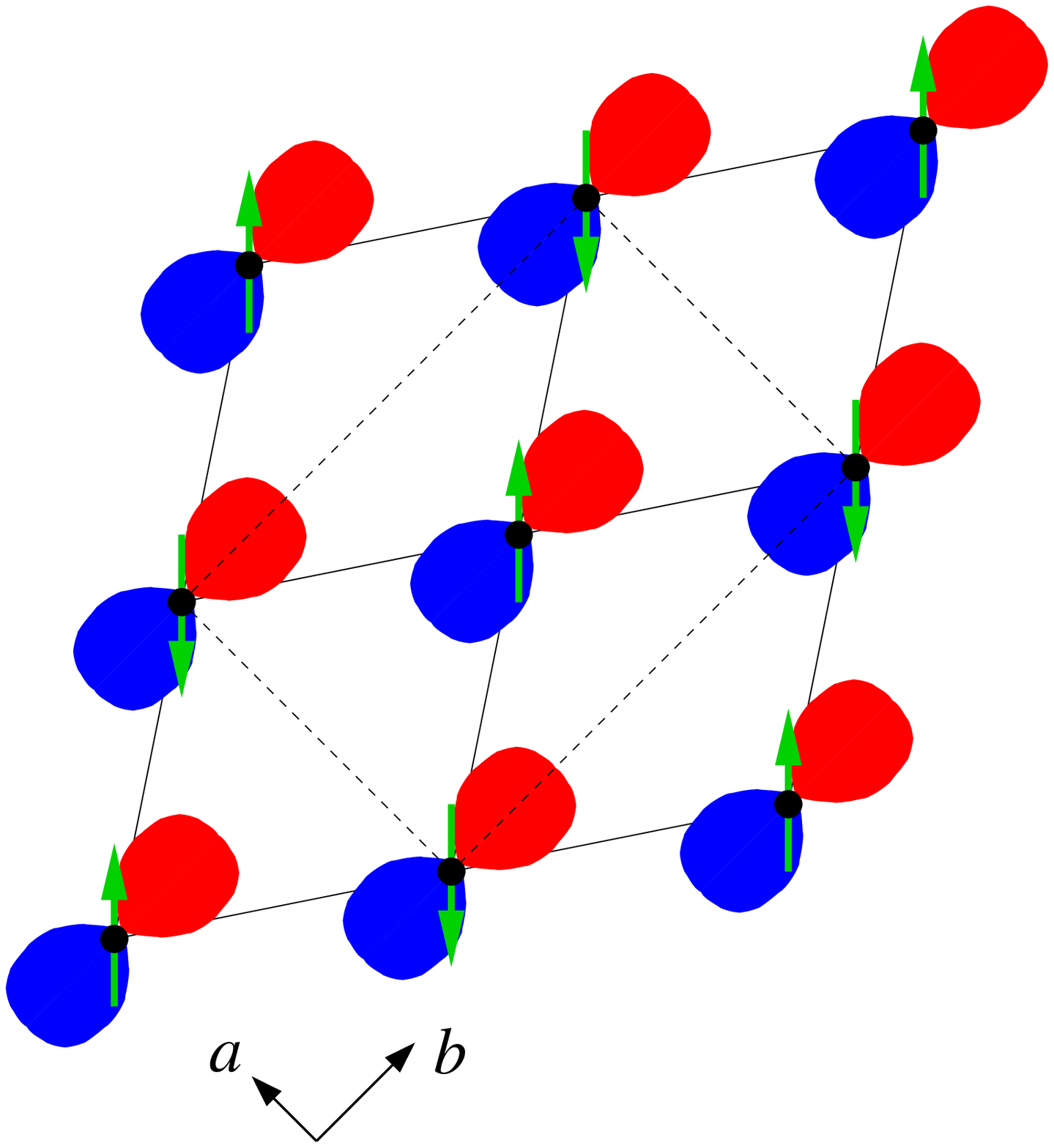}
  \includegraphics[width=0.42\linewidth]{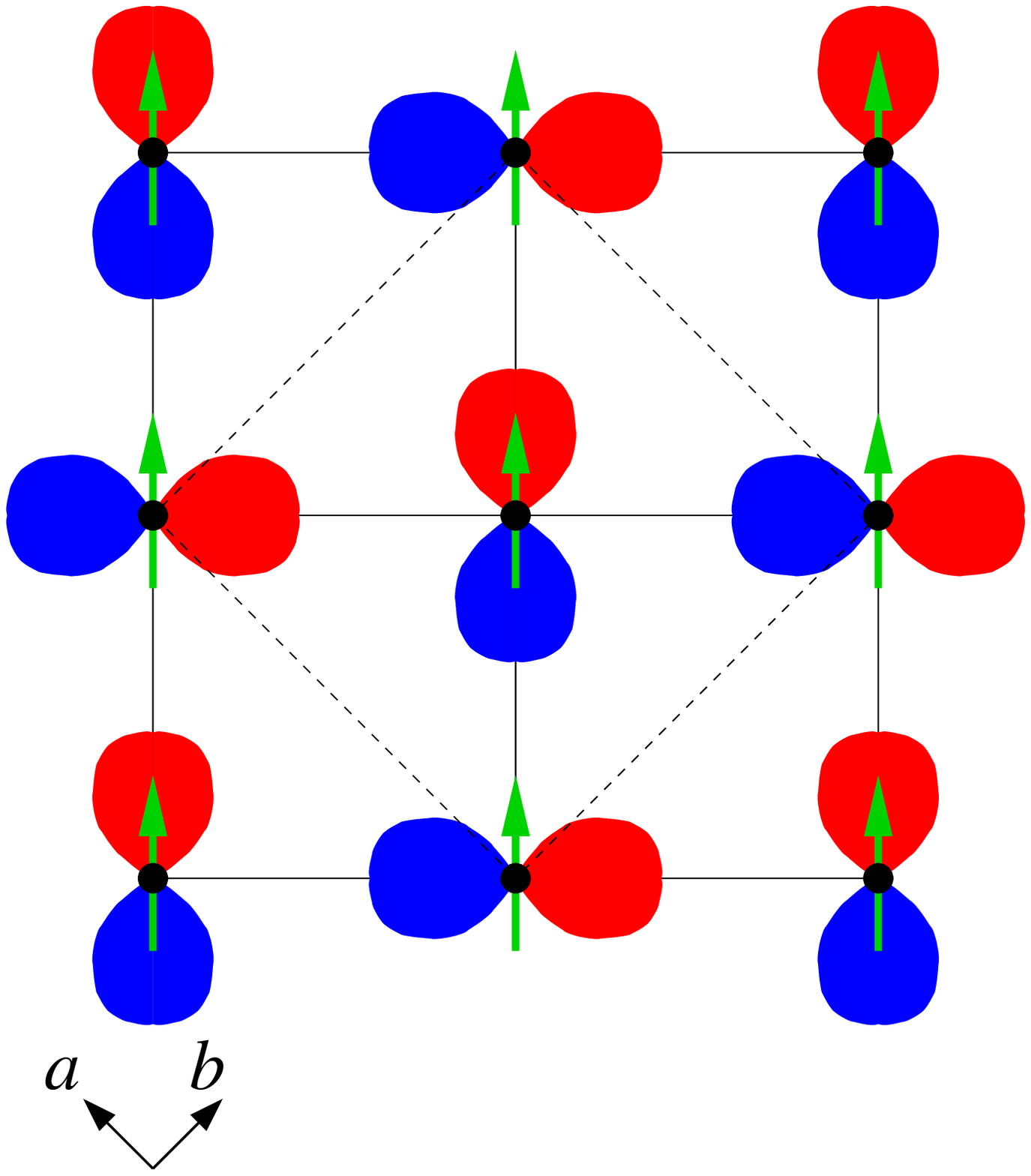}
\caption{Sketch of the low-temperature competing orders of the
system.  On the left hand side it is represented a lattice with
orthorhombic symmetry and ferro-orbital order plus
antiferromagnetism (FO \& AFM). On the right hand side, a lattice with
tetragonal symmetry with antiferro-orbital order plus
ferromagnetism (AFO \& FM).}
\label{fig:CompetingOrders}
\end{figure}
The competing low temperature orders are schematically shown in Fig.~\ref{fig:CompetingOrders}. For small values of $G$, the free energy minimization gives a ground state
with a strong lattice distortion breaking the tetragonal symmetry, as shown on the left
hand side of Fig.~\ref{fig:CompetingOrders}. In this case, the symmetry of the lattice is orthorhombic and ferro-orbital order (FO) coexists with antiferromagnetic order (AFM).
On the other hand, for large values of the elastic constant  $G$, anti-ferro orbital (AFO) and ferromagnetic (FM) spin order are simultaneously realized. This is sketched on the right hand side of Fig.~\ref{fig:CompetingOrders}.

\subsection{Absence of magnetic field.}
The $G$-$T$ phase diagram for $H=0$ is shown in Fig.~\ref{fig:phaseH0}. The elastic constant $G$ of the lattice controls the distortion. Apart from the high-temperature disordered phase (PO) we can distinguish two main regions in the phase diagram, as discussed below. 

\subsubsection{Soft lattice}
Lowering the temperature from the disordered phase we find for a soft lattice (small $G$) a crossover to a ferro-orbital ordered (FO) state. This crossover takes place at a temperature in the range of $k_BT_{FO}=\frac{K^2}{4G}$ indicated by the diffuse line in Fig.~\ref{fig:phaseH0}. The FO order is accompanied by a substantial orthorhombic distortion $\varepsilon\gg\varepsilon_0$ which is driven by the coupling of the strain field to the orbital degrees of freedom as described by Eq.~(\ref{eq:Hdis}). In addition, below a critical temperature $k_BT_{AFM}=J\left(B+A\eta_{\ba}^2\right)$, antiferromagnetic (AFM) order sets in.

\subsubsection{Hard lattice}
For a harder lattice (large $G$), the gain in energy by polarizing the orbital degrees of freedom is not sufficient to drive a substantial orthorhombic distortion and $\varepsilon\approx\varepsilon_0$. Instead, there is a staggered orbital order (AFO) which sets in at temperatures slightly below $k_BT_{AFO}=JC$. The AFO order drives a ferromagnetic (FM) ordered phase roughly below $k_BT_{FM}=-J\left[B+A\left(\eta_{\ba}^2-\frac{1}{4}\right)\right]$. For smaller values of $G$, this transition is weakly first-order but changes its character at the tricritical point (TP) to second-order.   

\subsubsection{Structural transition}
There is a structural transition between the soft and the hard lattice. Below $T_{AFM}$ this transition is of first-order and is characterized by a simultaneous discontinuity in all the order parameters. The first-order line splits into two second-order lines at a bicritical point (BP). For temperatures above $T_{AFM}$ the transition is characterized by the onset of a staggered orbital and a gradual suppression of the ferro-orbital order with a concurrent reduction of the lattice distortion. 

\begin{figure}
\includegraphics[width=1.0\linewidth]{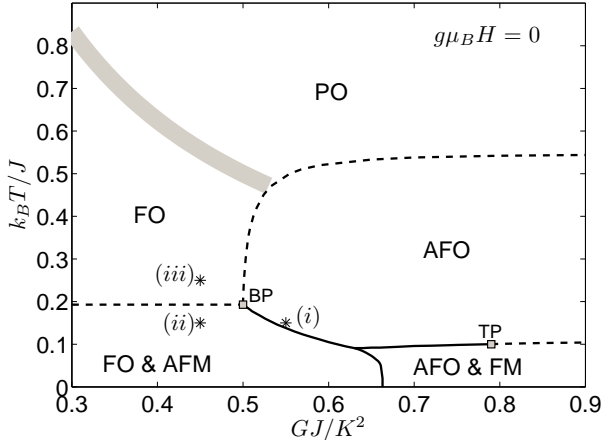}
\caption{$T-G$ phase diagram for $H=0$ involving antiferromagnetic (AFM), ferromagnetic (FM), ferroorbital (FO) and antiferroorbital (AFO) order. Dashed
lines indicate second order phase transitions while full lines represent first order transitions. TP is a tricritical point and BP is a bicritical point. The diffuse line between FO and paraorbital (PO) region indicates a crossover.} 
\label{fig:phaseH0}
\end{figure}

\subsection{Applied magnetic field}

\begin{figure}
\includegraphics[width=1.0\linewidth]{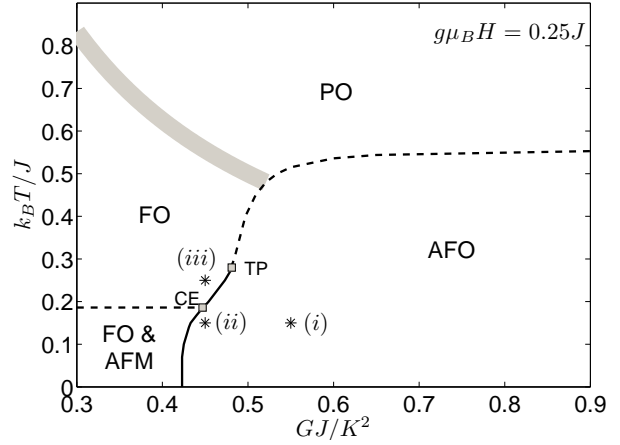}
\caption{$T-G$ phase diagram for $g\mu_BH=0.25J$. CE is a critical endpoint. The other abbreviations have the same meaning as in Fig.~\ref{fig:phaseH0}. } 
\label{fig:phaseH025}
\end{figure}

Now we introduce in our analysis the effect of an external
magnetic field applied in the $x$-direction, that enters in the
Hamiltonian by the term ${\cal H}_{mag}$ given in Eq.~(\ref{Hmag}). For $g\mu_BH=0.25J$ we obtain the phase diagram shown in Fig.~\ref{fig:phaseH025}. Comparing Fig.~\ref{fig:phaseH0}  and \ref{fig:phaseH025} we see that the main effect of the application of the magnetic field is the displacement of the first order structural transition towards smaller values of $G$ and, consequently, the reduction of the region of the phase diagram with AFM order, as expected. Therefore, the lattice effectively becomes harder in the presence of a finite magnetic field. The first order nature of the structural transition is now present for a larger range of temperatures, up to the tricritical point TP shown in Fig.~\ref{fig:phaseH025}. In addition, a critical end-point CE is now defined when the second order $k_BT_{AFM}$ line meets the first order structural transition line.

\subsection{Metamagnetic transition}
\begin{figure}
\centering
{\includegraphics[width=1\linewidth]{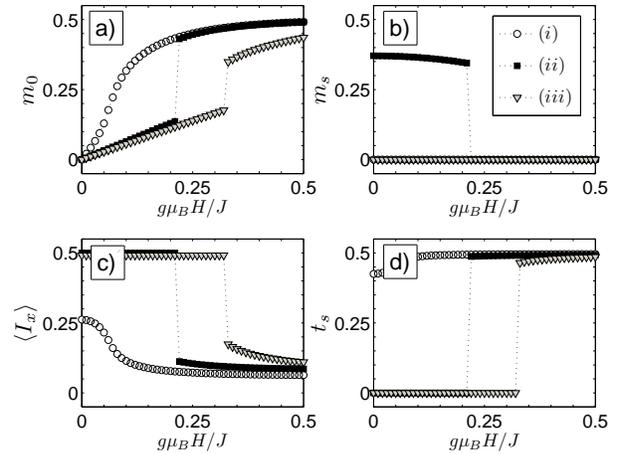}}
\caption{Evolution of the transverse magnetization $m_0$, the staggered magnetization $m_s$, the uniform orbital order $\langle I^x\rangle$ and the staggered orbital order parameter $t_s$
as a function of the applied magnetic field for three different sets of parameters $(k_BT/J,JG/K^2)$ given by $(i)$ (0.15,0.55), $(ii)$ (0.15,0.45) and $(iii)$ (0.25,0.45). The points $(i)-(iii)$ are also shown in the phase diagrams Figs.~\ref{fig:phaseH0} and \ref{fig:phaseH025}.} 
\label{fig:metamag}
\end{figure}
The results of our calculations include a first order metamagnetic transition (MMT). The characteristics of this transition are summarized in Fig.~\ref{fig:metamag}.
Here we show, from panels a) to d), how the magnetic and orbital order parameters change as a function of the
field strength, for different values of temperature $T$ and elasticity $G$. The different $(T,G)$ points are labeled by $(i)-(iii)$ and are also indicated in the phase diagrams Figs.~\ref{fig:phaseH0} and \ref{fig:phaseH025}. The curves $(ii)$ and $(iii)$ of Fig.~\ref{fig:metamag}a) show the discontinuous evolution of the magnetization towards a strongly polarized magnetic state by the application of a magnetic field. This magnetic
transition is accompanied in our system by a structural
transition where at the same
metamagnetic critical field $H_c$ an orthorhombic FO phase changes discontinuously
towards an AFO phase with tetragonal symmetry, as shown in Fig.~\ref{fig:metamag}c)-d). For the MMT to be observed it is not stringent that the zero-field phase has antiferromagnetic order, as long as it is close enough to such a phase. 
Notice however, that a MMT is only possible for a soft lattice (small $G$) at low temperatures, as it is the case for the $(T,G)$ points $(ii)$ and $(iii)$ of Fig.~\ref{fig:phaseH0}. For larger values of the elasticity, such as for $(i)$, we find a continuos evolution of the order parameters. 
In summary, for the critical field $H_c(T,G)$, we find the general behavior that both rising $T$ or lowering $G$ increases the critical field.

\begin{figure}
\centering
{\includegraphics[width=0.9\linewidth]{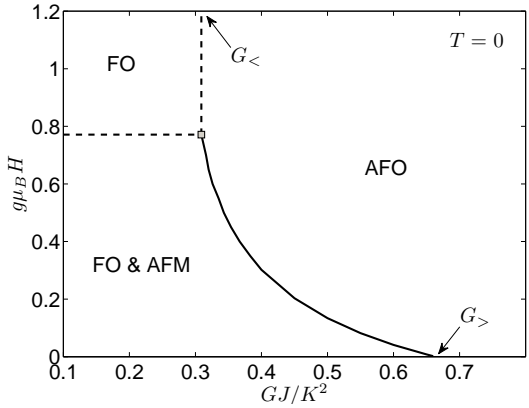}}
\caption{$H-G$ phase diagram for $T=0$. The values $G_<$ and $G_>$
bound the $G$-axis region where a metamagnetic transition is
allowed. The first order line (full) meets the second order line
(dashed) at a QCP.} \label{fig:phasediagramT0}
\end{figure}

Further insights concerning the conditions for the MMT to occur can be obtained from the $H$-$G$-phase
diagram at zero temperature shown in Fig.~\ref{fig:phasediagramT0}. It is worth noting that a first
order MMT (accompanied by a structural
transition) can only occur at $T=0$ if we apply a magnetic field
in the FO \& AFM zone, and only for elasticity values belonging to
the region $G_<<G<G_>$. For $\varepsilon_0\ll K/J$, $G_<$ and $G_>$ are given by
\begin{eqnarray}
G_<&=&\frac{K^2}{J(A+4C)-4K\varepsilon_0},\nonumber\\
G_>&=&\frac{K^2}{4J\left(A\left[1/4-2\eta_{\ba}^2\right]-2B+C-\frac{K\varepsilon_0}{J}\right)}.
\end{eqnarray}
Notice that at $G_>$ there is a first order structural phase
transition for $T,H=0$. In Fig.~\ref{fig:phasediagramT0} it can be
seen that the metamagnetic critical field decreases as $G$ is
strengthened. This qualitative behavior remains valid at finite but low temperatures. 

\subsection{Magnetostriction and thermal expansion}
\begin{figure}
 \centering
{\includegraphics[width=0.9\linewidth]{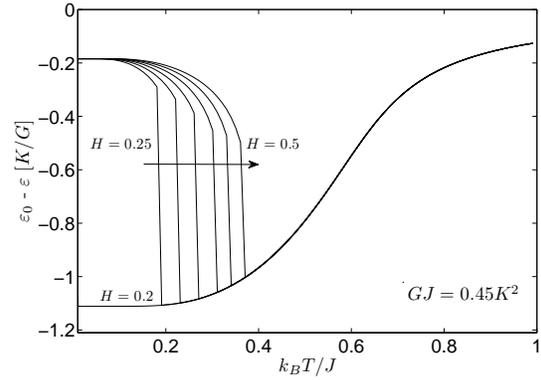}}
\caption{The orthorhombic distortion order $\varepsilon_0-\varepsilon = - \frac{K}{G}\langle I^x \rangle$ as function of the temperature for different values of the magnetic field between $H=0.2$ and $H=0.5$ (in units of $J/g\mu_B$).} \label{fig:epsilonT}
\end{figure}

Since there is a close relation between structural and metamagnetic transition the temperature dependence of the lattice parameters show a qualitative different low-temperature behavior for magnetic fields below and above the metamagnetic transition. In our model, changes of the lattice parameters are considered by the orthorhombic distortion field $\varepsilon$, related to the ferro-orbital order $\langle I^x\rangle$ by Eq.~(\ref{I^x}). An increase of the c-axis is assumed to be proportional to $\varepsilon_0-\varepsilon$. Therefore, we show in Fig.~\ref{fig:epsilonT} the temperature dependence of $\varepsilon_0-\varepsilon$ for different magnetic fields at $G=0.45K^2/J$. The $T=0$ critical field for this elasticity corresponds to $H_c\approx 0.22J/g\mu_B$ as it can be deduced, for example, from the zero temperature $G-H$ phase diagram of Fig.~\ref{fig:phasediagramT0}. For fields lower than $H_c$, a metamagnetic transition will never be reached and the heating of the system by increasing the temperature drives the lattice towards a disordered PO phase through the crossover region (diffuse line in Fig.~\ref{fig:phaseH0} and Fig.~\ref{fig:phaseH025}). 

However, for fields $H>H_c$ the system is in the metamagnetic region. There is a low temperature AFO order and consequently, the distortion $\varepsilon$ is small. By heating the system, metamagnetism is destroyed and at the same time the lattice undergoes a first order structural transition. This is seen in the $\varepsilon$ vs. $T$ plot (Fig.~\ref{fig:epsilonT}) by a jump of the distortion to a large negative value of $\varepsilon_0-\varepsilon$. If we keep heating the system, we reach again the PO region by passing the crossover zone. Obviously, the temperature of the 1$^{st}$-order AFO/FO transition is larger for higher magnetic fields. This is reflected in the evolution of the $\varepsilon_0-\varepsilon$ discontinuity in Fig.~\ref{fig:epsilonT} from $H=0.25J/g\mu_B$ to $0.5J/g\mu_B$.

\section{Comparison to experiments}
\label{sec:exp}
Eventually, we will motivate our microscopic model in view of the  experimental results. In particular, we will consider  the phase diagram at zero-field as well as at the magnetostriction and magnetization
measurements that characterize the metamagnetic transition in \CS.

\subsection{Phase diagram}

The theoretical zero-field phase diagram
(Fig.\ref{fig:phaseH0}) can be compared to the
experimental $T$-$x$-phase diagram obtained in Ref.~\cite{NM00} and schematized in Fig.~\ref{fig:PD}. We relate the phases on the left-hand side of Fig.~\ref{fig:phaseH0} (soft lattice) to the phases
that are observed in region II of the experimental phase diagram. In both cases the symmetry of the lattice is reduced. The tilting of the
RuO$_6$-octahedra observed in the real material is modeled by a distortion of the
2D lattice where the orbital and magnetic modes live. This reduction of symmetry is characterized by the strain field $\varepsilon$ or the ferro-orbital order $\langle I^x \rangle$. For larger $G$ 
$\langle I^x \rangle$ is reduced and we find a transition to an AFO (predominantly) tetragonal phase where, in addition,
FM order sets in at small temperatures. This is in fact the characteristics found in
region III of the experimental phase diagram near $x=0.5$. Therefore, it is not unfounded to relate the elasticity of the lattice (as modeled by $G$) to the Sr concentration $x$: both control the amount of distortion in their respective phase diagrams. Note that a Curie-like behavior of the orbital order $\langle I^x \rangle \approx K\varepsilon/4k_BT$ is shown in Fig.~\ref{fig:epsilonT}, which accounts for the temperature dependence of experimentally measured anisotropy of the  spin susceptibility in the distorted (orthorhombic) region \cite{NM00,ST04}.

\subsection{Metamagnetic transition}

The metamagnetic transition shown in Fig.~\ref{fig:metamag} can be
related to the experimental MMT \cite{B05,B06}
in the following way: in zero magnetic field, at a
temperature and doping ($G$ in our language) leading us  into
the low-temperature FO zone of the phase diagram (region II of
Fig.\ref{fig:PD}), we find a large strain $\varepsilon \gg \varepsilon_0$ and a zero-component of the magnetization along the $x$-direction. If we now
turn on the transverse magnetic field, a finite component of
$m_0$ appears, although for small enough fields the strain is
still present in the system. For some critical field $H_{c}(T,G)$
we find a first order transition in the magnetization $m_0$ which
jumps discontinuously to some larger value, while the strain drops
simultaneously to $\varepsilon \sim \varepsilon_0$. The transcription of this to
the experiments is that the octahedra returns to the structure it had
before tilting. Also the $c$-axes adapts to the initial
direction it had in region III of the phase diagram. The tetragonal symmetry needs, however,
not to be restored. This behavior explains now the reversal of the
structural distortion which occurs upon cooling at zero field, since applying a 
high magnetic field at low enough
temperatures leads back to the old structure, 
as shown in Fig.~\ref{fig:epsilonT} and seen experimentally \cite{B05}.

This first order transition in the magnetization corresponds to
the MMT observed in the experiments. Note that inhomogeneity 
is ignored in our description. What occurs as a discontinous first order transition here,
would be a smooth crossover (MMT) when disorder, e.g. in alloying Ca and Sr, is 
included  \cite{ST04}.

The critical elastic constants $ G_{<} $ and $ G_{>} $ found in
Fig.~\ref{fig:phasediagramT0}, bounding the segment on the $G$-axis
where the first order magnetic and structural transitions are
possible at $T=0$, can be mapped to the concentration
values $x$ of the experimental phase diagram that define the region
where the MMT can be observed. Therefore we may identify
$G_<$ with $x=0.2$ and $G_>$ with $x=0.5$. This relation is
consistent with the experimental results that show a smaller
energy scale for the transition at $x=0.5$ compare to the one at
$x=0.2$. The MMT is shifted towards lower
fields when the Sr content grows from $x=0.2$ to $0.5$. On
the other hand, the first order structural transition driven at
zero temperature for $G_>$ in the theoretical model can be related
to the structure quantum phase transition of \CS~ at $x=0.5$. At $G_>$, the metamagnetic transition may be considered as occurring at zero-temperature and zero magnetic field, in agreement with the experimental measurements that shows that the MMT seems to be shifted towards a field close to zero for $x=0.5$.

\subsection{Magnetostriction and thermal expansion}

The dependence of the first order magnetic transition on the
temperature shown in Fig.~\ref{fig:metamag} for points $(ii)$ and $(iii)$ of the phase diagram, and the temperature dependence of the distortion of Fig.~\ref{fig:epsilonT} can be compared to the
experimental measurements, too. This is actually the expected
behavior and reproduces some of the results shown in
Ref.~\cite{B05,B06}. If we look for example at the magnetostriction measurements of
Ref.~\cite{B05,B06}, where they show $\Delta L(H)/L_0$ along the $c$
axis as a function of the applied magnetic field, these results
can be interpreted as the response of the lattice structure to the
metamagnetic transition.  The jump of the magnetization is coupled
to an increase of the $c$-lattice constant $L_0$, since the structure of the lattice before the octahedra
tilting is restored  (see Fig.~\ref{fig:metamag}c)).

In addition, the thermal expansion coefficients and the integrated length changes measured below and above the critical field show that the pronounced shrinking of the octahedra  along the $c$-direction in zero field is successively suppressed by the field and turns into a low-temperature elongation at fields larger than $H_c$~\cite{B05,B06}. The experimental temperature dependence of $\Delta L/L_0$ for various magnetic fields, as a measure of the lattice distortion, follows a temperature dependence of $\varepsilon$ similar to the case of fields above and below $H_c$ and shown in Fig.~\ref{fig:epsilonT}. In fact, for fields below the MMT the lattice evolves from the low-temperature FO order to the high-temperature PO region. On the other hand, for fields above the MMT the system shows a tetragonal AFO symmetry within the MMT- region which is suppressed by increasing the temperature, switching to an orthorhombic FO order. For higher temperatures the system looses  orbital order and reaches the disordered PO region.

\section{Conclusion}
\label{sec:con}
In summary, we have analyzed a microscopic model for the description of
magnetic and structural properties in \CS which is based on the assumption that
two of the three electron bands known in Sr$_2$RuO$_4$ are Mott localized in 
regions II and Ca-rich zone of region III
(near $x=0.5$) of the phase diagram  \cite{ANKRS02}. The mean-field treatment of this model reproduces the basic magnetic and structural properties, as well as
the metamagnetic transition of this material. The elastic properties have been introduced assuming
that the elastic constant depends on the Ca-concentration. In this way we draw the connection
between our phase diagrams based on model parameters and the physical phase diagram
and find a good qualitative and in parts quantitative agreement. Our most important result is
the magnetostriction effect in connection with the metamagnetic transition which agrees well
on a qualitative level with recent experimental findings.

R.R. thanks M.P. L\'opez-Sancho for many useful discussions. R.R. acknowledges the hospitality of the ETH-Z\"urich, where part of this work has been done. This study was financially supported by the Swiss National fonds through the NCCR MaNEP and by the Center for Theoretical Studies of ETH Zurich. RR acknowledges financial support from MCyT (Spain) through grant FIS2005-05478-C02-01 and Agence Nationale de la Recherche Grant ANR-06-NANO-019-03.

\end{document}